\documentclass[
]{ceurart}

\usepackage{listings}
\begin{document}

\copyrightyear{2023}
\copyrightclause{Copyright for this paper by its authors.
  Use permitted under Creative Commons License Attribution 4.0
  International (CC BY 4.0).}

\conference{
Internation workshop on Knowledge Graph Generation from Text (TEXT2KG) Co-located with the Extended Semantic Web Conference (ESWC 2023) Date: May 28 - June 1, 2023, in Hersonissos, Greece}

\title{Science and Technology Ontology: A Taxonomy of Emerging Topics}

\tnotemark[1]
\tnotetext[1]{}


\author[1]{Mahender Kumar}[
orcid=0000-0001-7082-0475,
email=mahender.kumar@warwick.ac.uk,
]
\cormark[1]

\address[1]{Secure Cyber Systems Research Group, WMG, University of Warwick,
  Coventry, United Kingdom, CV4 7AL}

\author[1]{Ruby Rani}[
email=ruby.rani@warwick.ac.uk,
]

\author[1]{Mirko Bottarelli}[
email=mirko.bottarelli@warwick.ac.uk,
]

\author[1]{Gregory Epiphaniou}[
email=gregory.epiphaniou@warwick.ac.uk,
]

\author[1]{Carsten Maple}[
email=CM@warwick.ac.uk,
]


\begin{abstract} 
Ontologies play a critical role in Semantic Web technologies by providing a structured and standardized way to represent knowledge and enabling machines to understand the meaning of data. Several taxonomies and ontologies have been generated, but individuals target one domain, and only some of those have been found expensive in time and manual effort. Also, they need more coverage of unconventional topics representing a more holistic and comprehensive view of the knowledge landscape and interdisciplinary collaborations. Thus, there needs to be an ontology covering Science and Technology and facilitate multidisciplinary research by connecting topics from different fields and domains that may be related or have commonalities. To address these issues, we present an automatic Science and Technology Ontology (S\&TO) that covers unconventional topics in different science and technology domains. The proposed S\&TO can promote the discovery of new research areas and collaborations across disciplines. The ontology is constructed by applying BERTopic to a dataset of 393,991 scientific articles collected from Semantic Scholar from October 2021 to August 2022, covering four fields of science. Currently, S\&TO includes 5,153 topics and 13,155 semantic relations. S\&TO model can be updated by running BERTopic on more recent datasets.
\end{abstract}


\maketitle

\section{Introduction}

    Ontologies are a valuable tool for representing and organising knowledge about a specific topic or set of topics, using a set of concepts, relationships, and rules within the domain \cite{saif2012semantic,osborne2016automatic}. They have many applications, including data annotation and visualisation \cite{dudavs2018ontology}, forecasting new research areas \cite{salatino2018augur}, and scholarly data discovery \cite{fathalla2018semsur}. Some topic ontologies created in different domains include ACM Computing Classification System \footnote{The ACM Computing Classification System: http://www.acm.org/publications/class-2012.}, Physics and Astronomy Classification Scheme (PACS) \footnote{Physics and Astronomy Classification Scheme: https://publishing.aip.org/publishing/pacs}, replaced in 2016 by the Physics Subject Headings (PhySH) \footnote{PhySH - Physics Subject Headings: https://physh.aps.org/about.}, Mathematics Subject Classification (MSC) \footnote{2010 Mathematics Subject Classification: https://mathscinet.ams.org/msc/msc2010.html.}, the taxonomy used in the field of Mathematics, and Medical Subject Heading (MeSH) \footnote{MeSH - Medical Subject Headings: https://www.nlm.nih.gov/mesh.}. Creating these large-scale taxonomies is a complex and costly process that often requires the expertise of multiple domain experts, making it a time-consuming and resource-intensive endeavour. Consequently, these taxonomies are often difficult to update and maintain, quickly becoming outdated as new information and discoveries emerge. As a result, the practicality and usefulness of these taxonomies are significantly limited. One of the most notable advancements in ontology generation is the development of a large-scale automated ontology known as Computer Science Ontology (CSO) \cite{salatino2018computer}. CSO ontology defines a significant breakthrough in the representation of research topics in the computer science domain, providing a structured and comprehensive framework for organising and integrating knowledge but limited to computer science concepts only. 

  \textit{Research Challenge}: Understanding the dynamics associated with unconventional topics, which present a more comprehensive and holistic perspective of the knowledge landscape and interdisciplinary collaborations, poses a considerable challenge. Constructing an ontology for such unconventional topics necessitates recognising and collecting essential concepts and relationships from multiple domains.  Furthermore, unconventional topics may necessitate multidisciplinary study, necessitating the integration of information from many fields. By overcoming these challenges, there is an opportunity for researchers and academicians to study new and developing areas of science and technology, as well as facilitate interdisciplinary collaboration across varied fields.

   \textit{Contribution}. This paper presents preliminary work to construct a S\&TO ontology that automatically generates a taxonomy of unconventional S\&T topics.  S\&TO ontology is built by applying BERTopic to a dataset of 393,991 scientific articles collected from Semantic Scholar from October 2021 to August 2022, covering four fields of science: computer science, physics, chemistry and Engineering. Currently, S\&TO includes 5,153 topics and 13,155 semantic relations. Unlike existing ontology, S\&TO ontology can provide many benefits for knowledge representation and discovery, facilitating interdisciplinary research and enabling dynamic updates. 

    \textit{Organisation}. The rest of the paper is organized as follows. Section 2 discusses the dataset and methods for constructing the proposed S\&TO. Section 3 gives the proposed S\&TO. The experimental results are discussed in section 4. Section 5 presents the applications and Usecases of S\&TO, and the limitations of the current version are discussed in Section 6. Finally, the conclusion is given in section 7.

\section{Data and Methods}

\subsection{Semantic Scholar}

Semantic Scholar has many academic publications from various fields, including medical sciences, agriculture, geoscience, biomedical literature, and computer science. We used the RESTful Semantic Scholar Academic Graph (S2AG) API to retrieve a sample of these articles \cite{wade2022semantic}. This API offers users on-demand knowledge of authors, papers, titles, citations, venues, and more. We obtained 393,991 Science and Technology articles from Semantic Scholar using the S2AG API. The API provides a dependable data source that allows users to link directly to the related page on semanticscholar.org, making it a convenient and accessible way to retrieve information about academic papers.

\subsection{Methodologies}

\begin{figure}[h]
  \centering
  \includegraphics[width=120mm,scale=0.5]{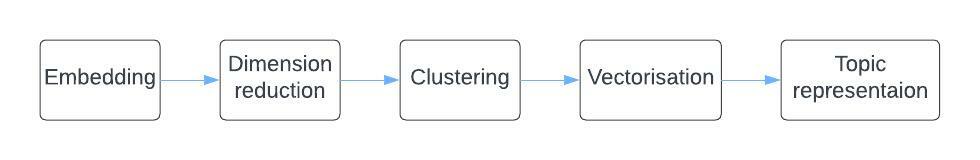}
  \caption{Data Flow}
  \label{Fig:flow}
\end{figure}

After downloading the dataset, we used the BERTopic method to obtain topics from the articles—some articles representing the multi-discipline need to be included as an outlier. To have these unconventional articles and reduce the outlier percentage, we adjusted parameters with BERTopic during topic clustering \cite{grootendorst2022bertopic}. Table 3 lists the critical BERTopic parameters used in the taxonomy generation.

As shown in Figure \ref{Fig:flow}, our suggested topic modelling workflow consists of five important steps: sentence embedding, dimension reduction, clustering, topic quality, and topic representation. The Sentence Embedding stage, in particular, involve turning textual input into numerical vectors that capture the underlying semantics of the text. Dimension Reduction is then applied to the vectors to lower their dimensionality and improve the effectiveness of the clustering procedure. Clustering is the process of combining similar vectors to generate coherent clusters of related material. The vectorisation ensures that the extracted topics are high quality, whilst topic Representation creates an interpretable summary of each topic. Overall, we provide a robust and effective approach to extracting meaningful unconventional topics from vast and heterogeneous datasets, utilising the power of BERTopic and careful parameter optimisation to assure optimal outcomes.

\begin{table}
  \caption{BERTopic Parameters}
  \label{tab:tab3}
  \begin{tabular}{rl}
    \toprule
    \textbf{Parameters} &	\textbf{Values}\\
    \midrule
    \textbf{Embedding parameter} \\
    \midrule
    embedding\_model &	sentence\_model\\

    \midrule
   \textbf{ UMAP parameters} \\
    \midrule
    n\_neighbors &	2 \\
    n\_components &	5 \\
    low\_memory &	true\\
    
    \midrule
    \textbf{HDBSCAN parameters} \\
    \midrule
    5min\_cluster\_size & 10\\
    metric	& euclidean\\
    cluster\_selection\_method &	eom\\
    prediction\_data	& true\\
    min\_samples	& 1\\

    \midrule
    \textbf{CountVectoriser parameters} \\
    \midrule
    Vocabulary &	Vocabulary \\
    Min\_df	& 10\\

    \midrule
    \textbf{c-TF-IDF parameters} \\
    \midrule
    top\_n\_words	& 30\\
    Verbose	&True\\
    min\_topic\_size &	20\\
    vectorizer\_model &	CountVectorizer\\
    low\_memory	& True\\
    calculate\_probabilities &	False\\
    Diversity &	0.4\\
    
  \bottomrule
\end{tabular}
\end{table}

\begin{table}
  \caption{BERTopic clustering model selection for outlier reduction}
  \label{tab:tab1}
  \begin{tabular}{cccc}
    \toprule
    Characteristics	& HDBSCAN  &	K-means	& HDBSCAN with \\
     	&  &	&  Probability\\
    \midrule
    Topic Quality &	Good &	Less &	Good \\
    Outlier &	High	& No &	Low \\
    Risk of missing  &	Less &	Little &	Least\\
    unconventional topics &	& &\\
  \bottomrule
\end{tabular}
\end{table}

\begin{enumerate}
    \item \textbf{Sentence Embedding}: We first transformed the input articles into numerical representations before analysing them. For this purpose, we utilised sentence transformers, the default embedding model used by BERTopic. This model can determine the semantic similarity of different documents. As default, BERTopic provides many pre-trained models among them we tried the following two: ``\verb|all-MiniLM-L6-v2|`` and ``\verb|paraphrase-MiniLM-L12-v2|``. While various sentence embedding models are available, we opted for the ``\verb|paraphrase-MiniLM-L12-v2|`` model in this work. This model effectively balances performance and speed, making it a good fit for our requirements. Thus, we can effectively translate textual data into numerical form and obtain relevant insights from large and diverse datasets using sentence transformers in conjunction with BERTopic.
    
    \item \textbf{Dimension Reduction}: Clustering can be complex since embeddings are often high-dimensional. To solve this problem, the dimensionality of the embeddings is frequently reduced to a more practical level. We used the UMAP (Uniform Manifold Approximation and Projection) technique, representing local and global high-dimensional features in a lower-dimensional domain \cite{mcinnes2018umap}. ``\verb|n_neighbors|`` and ``\verb|n_components|`` are two important parameters in the UMAP method. These parameters have a considerable impact on the size of the generated clusters. Larger values for these factors, in particular, result in the formation of more important clusters. We got optimal clustering results and extracted important topics from the input data by carefully tweaking these parameters.

    \item \textbf{Clustering}: BERTopic splits the input data into clusters of similar embeddings after the dimensionality reduction process. The clustering techniques' accuracy directly impacts the quality of the generated topics. K-means \cite{hartigan1979algorithm}, Hierarchical Density-Based Spatial Clustering of Applications with Noise (HDBSCAN) \cite{mcinnes2017HDBSCAN}, and Agglomerative Clustering \cite{mullner2011modern} are among the clustering techniques provided by BERTopic. The advantages and drawbacks of these clustering algorithms are summarised in Table \ref{tab:tab1}, emphasising their capacity to generate high-quality topics, manage outlier percentages, and limit the danger of missing unconventional topics. HDBSCAN is a density-based clustering algorithm used to find clusters of varying densities in a dataset. It works by constructing a hierarchical tree of clusters based on the density of the data points. It starts by identifying the points with the highest density and forming a cluster around them. Then, it gradually adds lower-density points to the cluster until a natural cutoff is reached, indicating the end of the cluster. According to our findings, the HDBSCAN with the prediction\_data parameter set to ``\verb|True|`` was the best option. Our method efficiently balances the above elements, allowing us to obtain meaningful and valuable insights from large and complex datasets.

    \item \textbf{Vectorisation}: The CountVectorizer technology turns text documents into vectors of phrase frequencies. However, it has significant drawbacks, such as failing to consider a specific topic's relative relevance in an article. To fix this issue, we adopted C-TF-IDF (Class-Based TF-IDF), a variant of the classic TF-IDF (Term Frequency-Inverse Document Frequency) method that allocates weights to terms depending on their relevance to a specific class of documents. We could fine-tune the model's performance in BERTopic by adjusting its parameters to optimise the clustering process using CountVectorizer with C-TF-IDF. This method made it possible to create higher-quality topics that are more fascinating and pertinent to the input data.

    \item \textbf{Topic Representation}: BERTopic can adjust TF-IDF to work at the cluster level instead of the document level to obtain a concrete representation of topics from the bag-of-words matrix. This modified TF-IDF is called c-TF-IDF. For word x in class c, the c-TF-IDF value is:

      \begin{equation}
          w_{x,c}=|tf_{x,c}| \times log(1+A/f_x)
      \end{equation}

    Where $tf_{x,c}$ denotes the frequency of word x in class c, $f_x$ denotes word x across all classes, and A denotes the average number of words per class.
\end{enumerate}

\begin{figure}[h]
  \centering
  \includegraphics[width=70mm,scale=0.5]{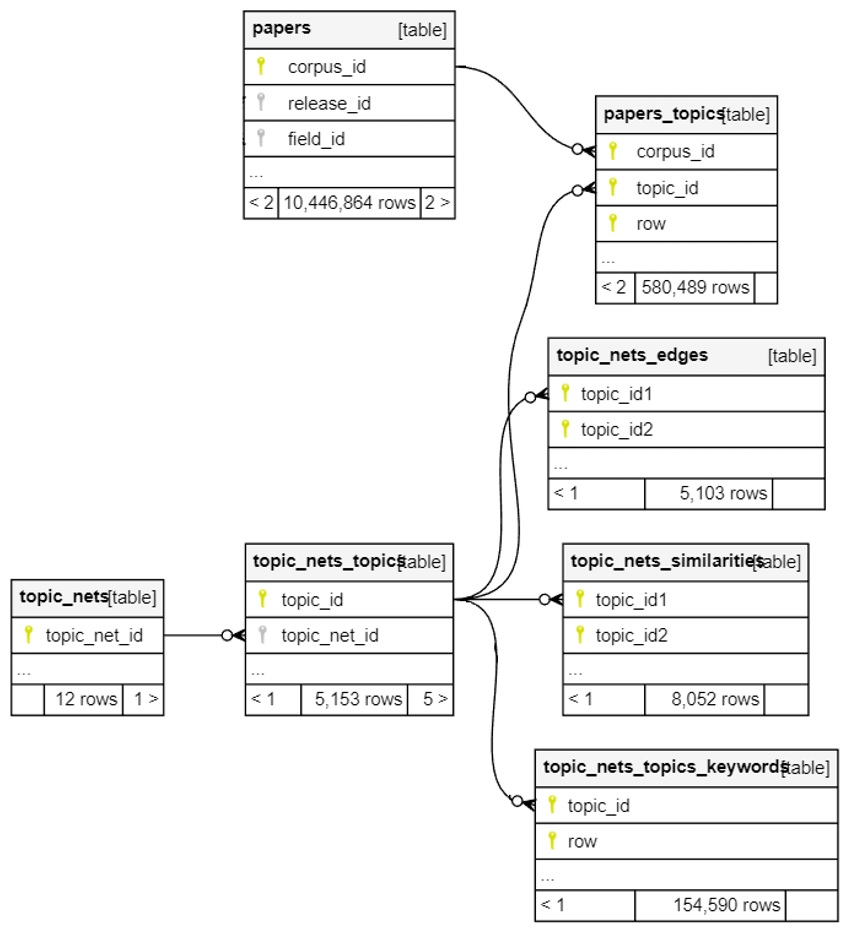}
  \caption{Schema of Topic Extraction and Topic Network Creation}
  \label{Fig:database}
\end{figure}

\section{Science and Technology Ontology Generation}

To create a topic network, also known as a knowledge graph, the metadata provided by the semantic scholar is utilised. The construction of topic ontologies involves the definitions of the following components:

\begin{itemize}
    \item \textit{Topics}: concepts of the topic ontology (e.g. Sports, Arts, Politics). 
    \item \textit{Predicates}: kinds of relationships that define the semantic link established between the ontology concepts. Many predicates can be defined in topic ontologies: hierarchical (e.g. superTopicOf) and non-hierarchical (e.g., part of, contribute to). 
    \item \textit{Relationships}: according to predicates and the set of elements they link, relationships are distinguished. They can be used to characterise the paths in the graphs and denoted as a triplet (T1, P, T2), where T1 and T2 denote the topics, and P denotes the predicate that links T1 and T2. 
\end{itemize}

\subsection{Topics}

The KeyBERT tool is used on the associated publications to extract keywords representing the essential concepts and topics within each document to produce the vocabulary for BERTopic. These keywords are then sent into BERTopic, which generates a complete collection of topics that capture the overarching topic found in the dataset. The extracted topics are saved in a database's "topic\_nets\_topics" and "topic\_nets\_topics keywords" (see Figure \ref{Fig:database}). Each topic's weight denotes the number of papers for which it serves as the main association, showing its relevance within the dataset.

There are two techniques to establish the relationship between papers and topics:

\begin{enumerate}
    \item \textit{Probability}: First, each paper's BERTopic/HDBSCAN probabilities are saved as entries in the "papers topics" table. These probabilities indicate how closely each document relates to each extracted topic.
    \item \textit{Embedding similarity}:  Second, using the "main topic id" field, the major topic associated with each paper, as identified by BERTopic, is directly linked to the paper using a SQL trigger. This allows for efficient querying and analysis of the topics and papers related to them within the corpus.
\end{enumerate}

\begin{figure}[h]
  \centering
  \includegraphics[width=120mm,scale=0.7]{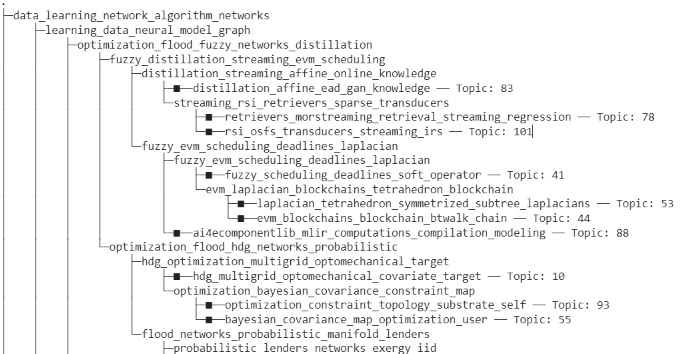}
  \caption{An instance of topic hierarchy}
  \label{Fig:Topic_hierarchy}
\end{figure}

\subsection{Relationship}
The next step is to create topic networks using the relationship among topics. Currently, the S\&TO ontology is built on 393,991 scientific papers collected from Semantic Scholar from October 2021 to August 2022. It covers four science fields: computer science, physics, chemistry and Engineering. S\&TO ontology follows the data model SKOS\footnote{SKOS Simple Knowledge Organization System - http://www.w3.org/2004/02/skos.} and includes the following semantic relationships:

\begin{itemize}
    \item ``\verb|relatedIdentical|``, It is a sub-metric of skos related, denotes that two topics can be viewed as identical for assessing research topics. The similarity between topics is calculated as cosine-similarity in the SQL stored procedure create\_topic\_nets. The relationship between topics is established if the similarity threshold is above 0.9.

    \item ``\verb|superTopicOf|``: It is a sub-metric of  skos:narrower, which means that a topic is a super-area of another topic in the graph. For example, "streaminmg\_rsi\_retrieval\_streaming\_regression" is the super-topic of Topics with topic\_ids 78 and 101, as shown in Figure \ref{Fig:Topic_hierarchy}.
    
    \item ``\verb|CommonArticles|``: It extracts common articles that appear in the two topics. The link between two topics is evaluated as the sum of the probability distribution by common articles assigned to the topics.

    \item ``\verb|nSimilarTopics|``: It returns the top x number of similar topics for an input keyword. For instance, the top 5 similar  topics related to the keyword "motor" are shown in Table \ref{tab:tab2}.
\end{itemize}

\begin{table}
  \caption{Top 5 similar topics to keyword "motor"}
  \label{tab:tab2}
  \begin{tabular}{cc}
    \toprule
    Topic Id	& Name\\
    \midrule
    33826	& 33826\_motorized\_spindle\_aerostatic\_restrictors\\
    50435	& 50435\_multimotor\_geartrain\_rocker\_truck\\
    62709 &	62709\_liner\_motorbike\_motored\_honing \\
    49677	& 49677\_motorized\_spindle\_nanocatalysts\_dragging \\
    46619	& 46619\_powertrains\_powertrain\_earthmoving\_2025\\
  \bottomrule
\end{tabular}
\end{table}.

\section{Experimental Results and Discussion}

In the literature \cite{fernandez2009makes}, ontology has been evaluated by four methods: gold-standard based \cite{maedche2002measuring}, corpus-based \cite{brewster2004data}, application-based \cite{sabou2007evaluating}, and structure-based methods \cite{buitelaar2004dynamic}. The gold-standard-based method compares the developed ontology with the referenced ontology developed earlier. The corpus-based method compares the significantly developed ontology with the contents of a text corpus that covers a given domain. The application-based approach considers applications and evaluations according to their performance across use cases. The structure-based approach quantifies structure-based properties such as size and ontology complexity.

Selecting the best evaluation approach and defining the rationale behind evaluating a developed ontology is necessary. In the proposed study, the science and technology ontology is in the early stages of development and will grow in future work. Thus, the application-based approach should not be a good evaluation approach because the proposed ontology is not proper for application purposes as it is currently in development. The proposed ontology is developed on a Semantic Scholar data set subset. Thus, the best reference ontology should be a semantic scholar. However, using Semantic Scholar as the gold standard ontology is impractical due to its unavailability.

\begin{figure*}
  \centering
  \includegraphics[width=150mm,scale=0.7]{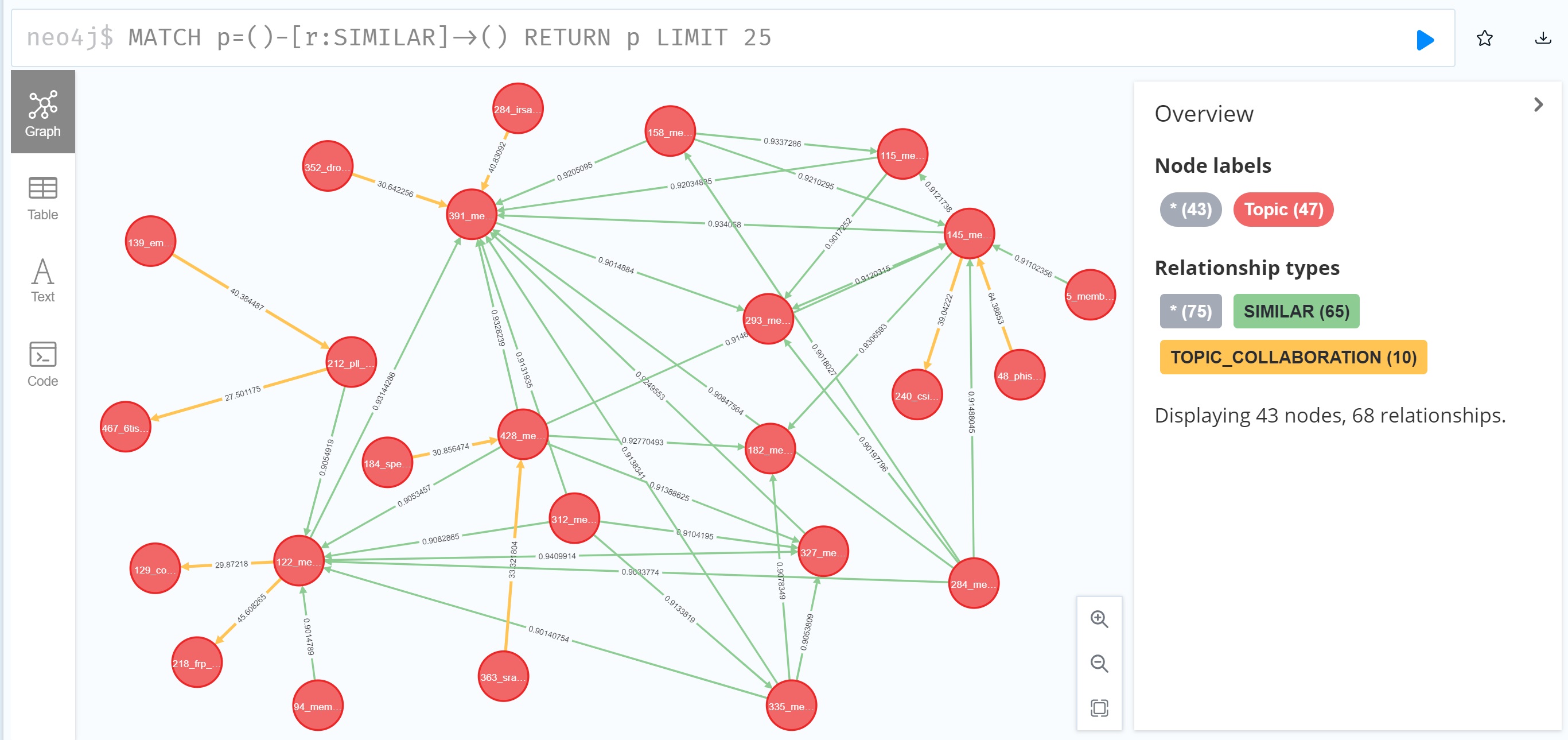}
  \caption{An instance of knowledge graph in Neo4j Browser \cite{fernandes2018graph}}
  \label{Fig:knowledge}
\end{figure*}

Structure-based evaluation is performed on several measures, including knowledge coverage and popularity measures (i.e., number of properties and classes) and structural measures (i.e., maximum depth, minimum depth, and depth variance). These measures are adopted based on the belief that densely populated ontologies with high depth and breadth variance are more likely to result in meaningful semantic content. Structural metrics are related to the semantic accuracy of adaptively modelled knowledge in the ontology \cite{sanchez2015semantic}.

In the context of the proposed S\&TO ontology, we quantified some structural measures by considering their taxonomic structure. S\&TO ontology gives 5,153 topics and 13,155 semantic relations (of which 8052 topics are based on cosine similarity and 5103 topics are based on probability distribution). Figure \ref{Fig:knowledge} shows an example of a knowledge graph covering topics and semantic relationships. We used neo4j as the graph database to host the final ontology \cite{fernandes2018graph}. Links in green indicate semantic relationships assessed using cosine similarity, while links in yellow indicate relationships based on the probability distribution of papers assigned to topics. S\&TO covers the maximum amount of articles for topic clustering and gives only 15.47\% of articles as an outlier, enabling the extraction of topics belonging to unconventional articles.

\subsection{Topics Details}

This section discusses the structure of topic networks, topics in topic networks and keywords related to topics in the network. 

\begin{table*}
  \caption{Structure of Table 'topic\_nets'}
  \label{tab:tab4}
  \begin{tabular}{cc}
    \toprule
    Field &	Description \\
    \midrule
    topic\_net\_id &	 	PK\footnote{Primary Key}, network unique identifier \\
    created\_on &		Indicates when the network was created \\
    status &	Indicates the status of the network (NEW, DONE, etc) \\
    year\_month &		Indicate the month from which this network was created \\
  \bottomrule
\end{tabular}
\end{table*}

The "topic\_nets" table (Table \ref{tab:tab4}) gives information about the development and status of topic modelling networks. Each network is assigned a unique identifier known as a "topic\_net\_id," its primary key. The "created\_on" parameter specifies the date and time the network was established. The "status" field offers information on the network's current status, which might take various values. This field assists in tracking the process's progress and ensuring that all networks are correctly generated and assessed. In addition, the "year\_month" parameter provides the month the network was founded. This feature is beneficial for tracking the temporal evolution of themes within the corpus since it allows researchers to understand how topics and their associations change over time.

\begin{table*}
  \caption{Structure of Table 'topic\_nets\_topics'}
  \label{tab:tab5}
  \begin{tabular}{cc}
    \toprule
    Field &		Description \\
    \midrule
    topic\_id &		PK, unique identifier of the topic\\
    topic\_net\_id	&	FK\footnote{Foreign Key} to the topic\_nets table
number	Integer	Topic number \\
    label &	Topic label\\
    topic\_weight &	Number of papers associated with this topic\\
    embedding &		Topic embedding used for cosine similarity\\
    similar\_topics &		Array of topic ids related to similar topics \\
  \bottomrule
\end{tabular}
\end{table*}

The "topic\_nets\_topics" table (Table \ref{tab:tab5}) provides essential information about the topics associated with the Topic networks. Each topic has a unique identifier known as a "topic\_id", the primary key. Each topic links to the corresponding network the "topic\_nets" using a unique identifier known as a "topic\_net\_id", its foreign key. A descriptive label is assigned to the topic based on the most common terms in the associated papers. The topic has "topic\_weight" which indicates the number of papers related to it, indicating its importance and relevance within the corpus. It also stores the "embedding", which will be used for cosine similarity calculations, and "similar\_topics", an array of topic ids related to searched topics.

\begin{table*}
  \caption{Structure of Table 'topic\_nets\_topics\_keywords'}
  \label{tab:tab6}
  \begin{tabular}{cc}
    \toprule
    Field &	Description \\
    \midrule
    topic\_id &		PK, unique identifier of the topic \\
    number &		Topic number\\
    row &		Auto-incremented number, ordered by increasing score\\
    keyword &		The keyword\\
    score &		The score associated with the keyword for this topic\\
  \bottomrule
\end{tabular}
\end{table*}

The "topic\_nets\_topics\_keyword" table (Table \ref{tab:tab6}) provides essential information about the keywords associated with the topic, represented by a unique identifier known as a "topic\_id", which is the primary key. It stores the fields such as: "number" representing the topic number, "row" is an auto-incremented number, and "keyword" representing the name of the keywords. In addition, "score" represents a weight associated with the keyword for that topic.

The "papers\_topics" table (table \ref{tab:tab7}) illustrates the  relationship between academic papers and the topics they cover. The "corpus\_id" denotes the unique identifier of the corpus the papers that were extracted from. The "topic\_id" represents the unique identifier of the topic that the paper covers, making it easier to track papers within that topic. The "probability" represents the weight of the paper assigned to the topic. This weight indicates the degree to which the paper covers the topic. The probability value is typically normalized, which is scaled to a range between 0 and 1.

\begin{table*}
  \caption{Structure of Table 'papers\_topics'}
  \label{tab:tab7}
  \begin{tabular}{cc}
    \toprule
    Field &		Description \\
    \midrule
    corpus\_id &	Part of PK, and FK to the papers table \\
    topic\_id &	Part of PK, and FK to the topic\_nets\_topics table \\
    row &		Auto-incremented number, ordered by increasing probability \\
    probability &		Probability of the paper to be assigned to this topic\\
  \bottomrule
\end{tabular}
\end{table*}

\subsection{Relation Details}

This section discusses the structure of topic relations, such as information on edges and similarities among topics. 

The "topic\_nets\_topics\_edges" stores (Table \ref{tab:tab8}) the information of edges among the topics "topic\_nets\_topics" in the network "topic\_nets". The edge is established between the two topics represented by their unique identifier (i.e., "topic\_id1" and "topic\_id2"). "edge\_Weight" is the sum of possibilities ("possibility" field of Table \ref{tab:tab7}) of papers sharing between two topics. The strength of collaboration "str\_of\_col" represents the weight computed as harmonic mean and normalised based on topics weights, shown in Eq (2). 

    \begin{equation}
          str\_of\_col = HarmonicMean\bigg(\frac{edge\_weight(id1,id2)}{topic\_weight(id1)} ,\frac{edge\_weight(id1,id2)}{topic\_weight(id2)}\bigg)
      \end{equation}

The "topic\_nets\_topics\_similarity" stores (Table \ref{tab:tab9}) the information of edges among the topics based on the similarity.

\begin{table*}
  \caption{Structure of Table 'topic\_nets\_topics\_edges'}
  \label{tab:tab8}
  \begin{tabular}{cc}
    \toprule
    Field &	Description \\
    \midrule
    topic\_id1 & Part of PK, and FK to the topic\_nets\_topics table\\
    topic\_id2 &	Part of PK, and FK to the topic\_nets\_topics table \\
    edge\_weight	&	 Sum of probabilities of papers sharing the two topics \\
    str\_of\_col  &	Field weight computed as harmonic mean and normalised weight\\
  \bottomrule
\end{tabular}
\end{table*}

\begin{table*}
  \caption{Structure of Table 'topic\_nets\_topics\_similarities'}
  \label{tab:tab9}
  \begin{tabular}{cc}
    \toprule
    Field &		Description \\
    \midrule
    topic\_id1 & Part of PK, and FK to the topic\_nets\_topics table\\
    topic\_id1  &	Part of PK, and FK to the topic\_nets\_topics table \\
    similarity  &	Cosine similarity between the two topics’ embeddings\\
  \bottomrule
\end{tabular}
\end{table*}

\section{ Advantages and Use-cases}

The proposed S\&TO with unconventional topics could have the following advantages and Use cases. 

\subsection{Knowledge expansion}
 S\&TO can broaden the scope of knowledge representation beyond existing ontologies by incorporating previously unconsidered topics. Offering a more holistic and comprehensive view of the knowledge landscape can lead to new insights and discoveries. In medical research, unconventional topics like holistic therapies or mindfulness practice might be incorporated into the ontology to provide a more comprehensive view of the more extensive health and wellness landscape \cite{howard2019artificial}.

 \subsection{Interdisciplinary Collaboration}
Second, by connecting topics from diversified fields that may have commonalities or be connected, an unconventional topics ontology encourage interdisciplinary collaboration. This can encourage the discovery of new research areas and cross-disciplinary cooperation, leading to novel solutions to complicated issues. For example, an ontology incorporating computer science and psychology issues could make it easier for academics in both domains to collaborate on human-computer interaction or affective computing \cite{xu2019toward}.

\subsection{Scalability and Adaptability}
An unconventional topics ontology has the benefit of being easily updatable and adaptable to reflect the most contemporary developments and topics, resulting in a dynamic and flexible knowledge representation system. This capability is significant in fast-paced sectors like technology and healthcare, where new topics and concepts develop regularly. For example, an ontology that includes topics relating to emerging technologies such as artificial intelligence \cite{zhang2021study} or blockchain \cite{monrat2019survey} can be easily updated to include new concepts and trends.

\section{Limitations}

The current version of the proposed S\&TO ontology has the following limitations. 

\subsection{Limited dataset}
The current version of S\&TO ontology is built on the Semantic scholar dataset covering 393,991 S\&T articles from October 2021 to August 2022. However, it could be built on more datasets. 

\subsection{Topic labelling}
Since S\&TO ontology utilises BERTopic, an unsupervised topic computation library, ontology suffers from the consequences of unlabeled topics.     Due to a lack of labelled data, it may be challenging to determine the significance and relevance of unconventional topics and distinguish them from noise or irrelevant topics. This is incredibly challenging when working with massive, complicated datasets containing various topics.

\subsection{Domain coverage}
S\&TO can capture various research topics and domains by covering these four domains: computer science, physics, chemistry and Engineering. However, many other disciplines and subfields within science and technology are not yet included in S\&TO. For example, biology, environmental science, and neuroscience are all essential areas of research that could be integrated into an ontology to create a more comprehensive and multidisciplinary framework for understanding scientific research. Expanding the coverage of S\&TO to include additional domains would have several potential benefits.

\subsection{Topic quality}
While S\&TO depicts a significant effort towards organising and categorising S\&T topics, there is still room for improvement regarding the quality of the topics included in the ontology.

\section{Conclusion and Future Work}

S\&T Ontology, an automated ontology of science and technology that includes all scientific study topics, was introduced in this paper. We constructed an ontology encompassing four different science domains by utilising BERTopic on a collection of 393,991 scientific articles acquired from Semantic Scholar from October 2021 to August 2022. S\&TO can be updated using BERTopic on recent datasets, offering a dynamic and flexible foundation for knowledge representation. S\&TO ontology has the potential to broaden the scope of knowledge representation and stimulate interdisciplinary collaboration, making it a valuable resource for scientists and technologists.

The S\&TO ontology constantly evolves and requires ongoing enhancements to meet the expanding knowledge landscape's demands. Currently, S\&TO is growing, and we are employing topic labelling techniques to improve the organisation and comprehension of different topics by giving them meaningful "tags". This makes it easy for users to browse the ontology and derive valuable insights. Furthermore, we intend to improve the topic quality by investigating additional methodologies and algorithms for topic modelling and clustering. This will strengthen the ontology's accuracy and efficacy in describing the knowledge landscape. In addition, we intend to expand the ontology to a more extensive dataset, allowing for the inclusion of more unconventional categories and topics, which will improve and diversify the knowledge base.

\bibliography{sample-ceur}


\appendix

\section{BERTopic Parameters}
 Here, we summarised the parameters we set throughout S\&TO development. The following parameters have been set:

 \begin{itemize}
        \item ``\verb|n_neighbors|`` refers to the number of neighbouring data points needed to estimate the manifold. Large sample point embeddings produce a more global perspective of the structure, while low values produce a narrower one. To get a good strike, we set n=2 as the result of the estimation.
        \item ``\verb|n_components|`` refers to the number of components after the reduction in dominance. This value directly affects the clustering performance, so it is necessary to set an optimal value. By default, it is set to 5 to reduce the dimensionality as much as possible while maximizing the information in the generated embeddings.
        \item ``\verb|low_memory|``: It is set to TRUE because we use a huge dataset and need a lot of memory.

        \item ``\verb|min_cluster_size|``: The number of cluster generations highly relies on the cluster size. It is necessary to adjust the minimum size. After several experiments, a cluster size of 50 was found to be the optimal one. While high value gives few clusters of considerable size, and low value gives microclusters.
        \item ``\verb|metric|``: metric, like HDBSCAN, calculates the distances. Here, we went with Euclidean as, after reducing the dimensionality, we have low dimensional data, and not much optimisation is necessary. However, if you increase ``\verb|n_components|`` in UMAP, it would be advised to investigate metrics that work with high dimensional data.
        \item ``\verb|prediction_data|``: Make sure you always set this value to True, as it is needed to predict new points later. You can set this to False if you do not wish to predict any unseen data points.
        \item ``\verb|min_samples|``: It is automatically set to ``\verb|min_cluster_size|`` and controls the number of outliers generated. Setting this value significantly lower than ``\verb|min_cluster_size|`` might help you reduce the amount of noise you will get. Do note that outliers are typical to be expected, and forcing the output to have no outliers may not properly represent the data.
    
        \item ``\verb|top_n_words|`` refers to the number of words extracted per topic. In practice, we keep this value below 30, preferably between 10 and 20. The reasoning is that the more words representing a topic, the less relevant it may be. In this case, the top words are most representative of the topic and are the focus.
        \item ``\verb|min_topic_size|`` specifies the minimum size of a topic. The lower the size value, the more topics are created. If the value is set too high, no topics may be created. We set this value too low, and we get many micro-clusters.
        \item ``\verb|calculate_probabilities|``  give probabilities of all topics per document. This could slow down the extraction of topics for a large number of many documents. 
        \item ``\verb|low_memory|`` set to true ensures that less memory is used in the calculations. This slows computation but allows UMAP to run on machines with little memory.
        \item ``\verb|diversity|`` reports a range of topic diversity from 0 to 1, where 0 indicates no diversity and 1 indicates a lot of diversity. Higher diverse topics mean less coherent topics in smaller cluster sizes. In our case, the diversity is assumed to be 0.4 or above.
    \end{itemize}

\end{document}